\documentclass[10pt,conference]{IEEEtran} 
\IEEEoverridecommandlockouts
\usepackage{cite}
\usepackage{amsmath,amssymb,amsfonts}
\usepackage{algorithmic}
\usepackage{graphicx}
\usepackage{textcomp}
\usepackage{booktabs}
\usepackage{color}
\usepackage{algorithm}
\usepackage{diagbox}
\usepackage{url}
\usepackage{xcolor}
\usepackage{listings}
\usepackage{hyperref}
\usepackage[normalem]{ulem}
\useunder{\uline}{\ul}{}
\usepackage{multirow}
\def\BibTeX{{\rm B\kern-.05em{\sc i\kern-.025em b}\kern-.08em
    T\kern-.1667em\lower.7ex\hbox{E}\kern-.125emX}}

\newcommand{\code}[1]{\textcolor{gray}{\texttt{#1}}}

\IEEEoverridecommandlockouts
\usepackage{tikz}
\usepackage{textcomp}
\usepackage{hyperref}
\usepackage{lipsum}

\newcommand\copyrighttext{%
  \footnotesize \textcopyright 2023 IEEE. Personal use of this material is permitted.
  Permission from IEEE must be obtained for all other uses, in any current or future 
  media, including reprinting/republishing this material for advertising or promotional 
  purposes, creating new collective works, for resale or redistribution to servers or 
  lists, or reuse of any copyrighted component of this work in other works. 
  DOI: \href{https://doi.org/10.1109/MSR59073.2023.00034}{10.1109/MSR59073.2023.00034}}
\newcommand\copyrightnotice{%
\begin{tikzpicture}[remember picture,overlay]
\node[anchor=south,yshift=10pt] at (current page.south) {\fbox{\parbox{\dimexpr\textwidth-\fboxsep-\fboxrule\relax}{\copyrighttext}}};
\end{tikzpicture}%
}

\begin{document}

\title{Cross-Domain Evaluation of a Deep Learning-Based Type Inference System}

\author{\IEEEauthorblockN{1\textsuperscript{st} Bernd Gruner}
	\IEEEauthorblockA{\textit{Institute of Data Science} \\
		\textit{German Aerospace Center}\\
		Jena, Germany \\
		bernd.gruner@dlr.de}
	\and
	\IEEEauthorblockN{2\textsuperscript{nd} Tim Sonnekalb}
	\IEEEauthorblockA{\textit{Institute of Data Science} \\
		\textit{German Aerospace Center}\\
		Jena, Germany \\
		tim.sonnekalb@dlr.de}
	\and
	\IEEEauthorblockN{3\textsuperscript{rd} Thomas S. Heinze}
	\IEEEauthorblockA{\textit{Cooperative University} \\
		\textit{Gera-Eisenach}\\
		Gera, Germany \\
		thomas.heinze@dhge.de}
	\and
	\IEEEauthorblockN{4\textsuperscript{th} Clemens-Alexander Brust}
	\IEEEauthorblockA{\textit{Institute of Data Science} \\
		\textit{German Aerospace Center}\\
		Jena, Germany \\
		clemens-alexander.brust@dlr.de}
}

\maketitle
\copyrightnotice

\begin{abstract}
	Optional type annotations allow for enriching dynamic programming languages with static typing features like better Integrated Development Environment (IDE) support, more precise program analysis, and early detection and prevention of type-related runtime errors. Machine learning-based type inference promises interesting results for automating this task. However, the practical usage of such systems depends on their ability to generalize across different domains, as they are often applied outside their training domain.

	In this work, we investigate Type4Py as a representative of state-of-the-art deep learning-based type inference systems, by conducting extensive cross-domain experiments. Thereby, we address the following problems: class imbalances, out-of-vocabulary words, dataset shifts, and unknown classes.

	To perform such experiments, we use the datasets ManyTypes4Py and CrossDomainTypes4Py. The latter we introduce in this paper. Our dataset enables the evaluation of type inference systems in different domains of software projects and has over 1,000,000 type annotations mined on the platforms GitHub and Libraries. It consists of data from the two domains web development and scientific calculation.

	Through our experiments, we detect that the shifts in the dataset and the long-tailed distribution with many rare and unknown data types decrease the performance of the deep learning-based type inference system drastically. In this context, we test unsupervised domain adaptation methods and fine-tuning to overcome these issues. Moreover, we investigate the impact of out-of-vocabulary words.

\end{abstract}


\section{Introduction}

\indent Dynamically typed programming languages allow the annotation of optional data types by language extensions, like Python with PEP~484~\cite{Pep}, to compensate for their shortcomings \cite{Hanenberg2013AnES, TypeNotType}. Recent machine learning-based type inference approaches try to mitigate the drawbacks of static and dynamic approaches like imprecision due to applied abstraction or missing coverage~\cite{10.1145/2989225.2989226} and provide promising results \cite{Type4py, TypeBert, TypeWriter, Typilus}.

From other applications of machine learning for software engineering, such as vulnerability detection, several issues are already known, e.g., newly introduced vocabulary \cite{karampatsis2020big, hellendoorn2017deep}, an unbalanced class distribution \cite{imbclassvul}, and cross-domain predictions \cite{crossvul, imbclassvul}. Therefore, in this study, we investigate the deep learning-based type inference system Type4Py \cite{Type4py}, for potential problems that affect the prediction performance and limit the practical applicability of the system.

In our extensive experiments, we focus on exploring how cross-domain prediction and associated problems, such as dataset shifts, influence the results of the type inference system. Cross-domain means that the system is applied to data from domains other than the training data, which is the case in a real-world scenario. A domain represents a software area, such as web development or scientific calculation. These domains are determined by the Python Developers Survey \cite{JetBrainsSurvey}. We also address the problem of unknown classes, which can be caused by the cross-domain setting. The system is unaware of data types that are not present during training. If the data types from the target domain are not present in the training data, they cannot be predicted by the system. We investigate how these unknown data types affect the results of the type inference system.

Moreover, the distribution of data types is highly imbalanced, and we study the impact of this because the rare data types that have low support in the dataset are typically predicted less accurately by deep learning-based systems \cite{cao2019learning}. Furthermore, new vocabulary is introduced by the source code at a higher rate than in natural language \cite{hellendoorn2017deep}, resulting in more out-of-vocabulary words (OOV) that cannot be embedded by the Word2Vec approach \cite{Word2Vec}. We investigate how the resulting loss of information affects the recognition rate of the system.

In this paper, we demonstrate the negative impact of the mentioned issues on the Type4Py type inference system and aim to mitigate them by using transfer learning methods \cite{dann,wasserstein}. For our cross-domain experiments we use the benchmark dataset ManyTypes4Py \cite{mt4py2021} and our new CrossDomainTypes4Py dataset, which we present in this paper. Our dataset covers the two distinct, but most widely used code domains of web development and scientific calculation according to the Python Developers Survey \cite{JetBrainsSurvey}. It allows us to examine the differences between the domains and how they affect the performance of type inference systems.

In summary, we contribute the following:
\begin{description}
	\item[CrossDomainTypes4Py Dataset]\hfill \\
	Our dataset is publicly available\footnote{\url{https://zenodo.org/record/5747024}} and contains 7,912 repositories from the scientific calculation and the web domain with 682,354 and 341,029 type annotations, respectively. For the preprocessed version of the data, we removed duplicate repositories and files. We split the remaining data into training, validation \& test, extract relevant information and prepare them as input for the Type4Py system (see Section \ref{subsec:preprocessing}).
	\item[Cross-domain Experiments with Type4Py]\hfill \\
	We perform extensive cross-domain experiments with the state-of-the-art type inference system Type4Py and provide a detailed evaluation (see Section \ref{sec:evaluation}). We investigate how well the system can generalize across domains, which problems occur, what has to be considered, and possible ways to mitigate these issues.

\end{description}

In order to ensure the reproducibility of our experiments, we make our mining and preprocessing scripts to create the dataset available\footnote{\url{https://gitlab.com/dlr-dw/type-inference}}, as well as our experimental pipeline. Furthermore, we provide a repository list of our dataset.

The paper is structured as follows. In Section \ref{sec:relatedWork}, a literature review on deep learning-based type inference systems and existing datasets is given. This is followed by Section \ref{sec:methods}, which explains the occurring problems and the Type4Py method. Section \ref{sec:dataset} describes the creation of the dataset including the preprocessing steps. We use Section \ref{sec:evaluation} to present our research questions, evaluate the experiments and afterward answer the research questions. In the succeeding Section \ref{sec:limitations}, the limitations of our approach are described. Finally, a summary with an outlook is given in Section \ref{sec:conclusion}.

\section{Related Work}
\label{sec:relatedWork}
The first part of the section contains an overview of deep learning-based type inference systems. In the second part, available datasets for deep learning-based type inference are presented.

\subsection{Deep Learning-based Type Inference Systems}
The majority of publications in the area of deep learning-based type inference address the programming languages JavaScript/TypeScript \cite{DeepTyper, NL2Type, LambdaNet, Opttyper, TypeBert} and Python. In this study, we focus on the latter.

One of the first deep learning-based type inference systems for Python is DLType \cite{Dltpy}, which is similar to the approach presented in \cite{NL2Type}. DLType additionally uses natural language elements of the code, like comments and identifier names to make type prediction more accurate. The network architecture is based on a Recurrent Neural Network (RNN). However, it can only predict the 1000 most frequent data types. Another method is PyInfer \cite{Pyinfer}, which uses additional code context and Byte Pair Encoding (BPE) \cite{BytePairEncoding}. The latter helps mitigate the out-of-vocabulary (OOV) problem. Again, the number of predictable data types is limited to 500. A further improvement in classification accuracy is achieved in the TypeWriter \cite{TypeWriter} approach by combining a probabilistic guessing component and a type checker that verifies the proposed annotations. The method is limited to the 1000 most frequent data types. Typilus~\cite{Typilus} addresses this problem, through the use of deep similarity learning, which makes it possible to predict user-defined and rare data types that occur in the training data. For feature generation, a Graph Convolutional Neural Network (GCNN) is used. A similar approach is presented in \cite{ivanov2021predicting}, where a combination of Graph Neural Network (GNN) and FastText \cite{FastText} embeddings is investigated. For the processing of the features, a Text Convolutional Network is applied.

The Type4Py \cite{Type4py} method uses hierarchical Long Short-Term Memories (LSTM) networks for feature extraction in combination with a deep similarity learning approach. Thus, all data types seen in the training can be predicted, similar to \cite{Typilus}. Another method is HiTyper \cite{Hityper}, which uses a staged approach of static inference and deep neural network prediction. The two approaches are used alternately and complement each other.

None of the previously mentioned papers conducts a cross-domain evaluation or investigates the OOV problem and its effects on the results of the system. Such studies are relevant to examine the performance of the systems when using them outside their trained domain, which is the case in practical applications. A related method \cite{Crosslingual} transfers the knowledge of a type inference system across programming languages, but the authors of this paper are faced with fundamental problems caused by the difference in programming languages. The focus is on this fundamental difference and how knowledge can be reused despite language-specific constructs and different type systems. However, different domains in the same language are not considered. The study is thus at a different level of detail compared to our work. This leads primarily to other problems as we have them, where ours reside on a more detailed level within the same language. Therefore, a direct comparison with the study is not meaningful.

To the best of the authors' knowledge, no previous study conducts a cross-domain evaluation of deep learning-based type inference systems and investigates the corresponding problems. We choose the Type4Py approach for our investigations because its source code is available and according to the evaluation by Mir et al. \cite{Type4py}, it outperforms other state-of-the-art deep learning-based type inference systems.

\subsection{Datasets}
\indent There are already some extensive Python corpora \cite{10.1145/3022671.2984041, 8816757, orru2015curated}. However, these were not created specifically for type inference and thus no focus was placed on whether the projects had type annotations. These are needed as ground truth data for supervised learning and evaluating the systems. Many projects do not have type annotations and are therefore unsuitable for this task.

\indent The authors of machine learning-based type inference methods for Python usually present their own datasets. These datasets have the following downsides: only partly publicly available \cite{TypeWriter}, not very comprehensive \cite{ivanov2021predicting, Typilus}, and partly designed for special preprocessing steps \cite{Typilus, Dltpy, ivanov2021predicting}. An example that does not come with these downsides is the large and publicly available ManyTypes4Py dataset \cite{mt4py2021}. However, for a cross-domain evaluation, several datasets containing various domains are required. Therefore, we present CrossDomainTypes4Py with two subsets from different domains.

\section{Theoretical Background}
\label{sec:methods}
In the first section, we explain the problems which are addressed in this paper. In the following part, we describe the structure and operation of the Type4Py system.

\subsection{Problem Definition}
\label{subsec:dataset_shift}
This section briefly outlines the three main problems examined in this paper.\\

\subsubsection{Out-of-vocabulary Words}
The out-of-vocabulary problem is a general issue when applying machine learning-based methods to source code \cite{karampatsis2020big}. The problem is particularly prominent in this area since source code introduces new vocabulary at a higher frequency than natural language~\cite{hellendoorn2017deep}. The new vocabulary is created by the programmer, for example, in the form of class and variable names. However, the embedding method Word2Vec \cite{Word2Vec} used in Type4Py can only embed known words. Therefore, new words outside the vocabulary (OOV) cannot be embedded, resulting in a loss of information. This can affect the performance of the type inference system.\\

\subsubsection{Unknown Classes and Imbalanced Distribution}
Unseen or unknown classes are classes that appear in the test but not in the training set and therefore cannot be predicted by the system. This may be due to the fact that the training and test sets are from different domains and do not share the same classes. Furthermore, for datasets with unbalanced class distribution, e.g., with a long-tailed distribution, it happens that there are few classes with a lot of support and many classes with little support in the dataset. Machine learning-based systems have a noticeably better recognition rate for classes with a lot of support than for classes with little support \cite{cao2019learning}. It is important to know the impact of these two aspects to estimate the performance of the system in a real-world scenario.\\

\subsubsection{Cross-Domain Prediction and Dataset Shift}
Dataset shift is an important topic in machine learning since many real-world applications are affected by shifts and this harms the performance of the systems \cite{finlayson2021clinician, janez2022review}. According to \cite{MorenoTorres2012AUV} a dataset shift appears when training and test joint distributions are different, which can be defined as follows:
\begin{equation}
	P_{train}(y,x) \neq P_{test}(y,x),
\end{equation}
where $x$ is a set of features or covariates, $y$ is a target variable, and $P(y,x)$ is a joint distribution.
The dataset shift is very general and includes all possible changes between training and test distribution. Covariate and prior probability shifts are examples of dataset shifts and describe differences in feature and class distributions, respectively. These harm the accuracy of the system and can occur when training and test data come from different datasets or domains.

\subsection{Type Inference System}
\label{subsec:type4py}
\begin{table}[]
	\centering
	\caption{Key characteristics from the CrossDomainTypes4Py dataset are displayed in this table, broken down by domain. Here cal stands for the scientific calculation subset.}
	\begin{tabular}{@{}llll@{}}
		\toprule
		Criterion                 & Total     & Cal       & Web       \\ \midrule
		Repositories              & 7,912     & 4,783     & 3,129     \\
		Total Files               & 8,580,167 & 6,103,661 & 2,476,506 \\
		Python Files              & 2,791,989 & 2,111,694 & 680,295   \\
		Files after Deduplication & 636,516   & 470,011   & 166,505   \\ \bottomrule
	\end{tabular}
	\label{tab:Dataset}
\end{table}

\begin{table*}[]
	\centering
	\caption{Excerpt of the repository list.}
	\label{tab:dslist}
	\begin{tabular}{@{}ll@{}}
		\toprule
		URL                                        & Commit Hash                              \\ \midrule
		\ldots{}                                                                              \\
		https://github.com/arXiv/arxiv-base.git    & b20db1f41731f841106a0b53fb64fc3faa056b4f \\
		https://github.com/Double327/CDCSonCNN.git & 77d28b074d67e9f96ffdfcb94e24762fbe749457 \\
		\ldots{}                                                                              \\ \bottomrule
	\end{tabular}
\end{table*}
We use the Type4Py system as the basis for our investigation. In this section we provide a brief overview, for more detailed information please refer to Mir et al.~\cite{Type4py}.

During the preprocessing, the Python source code files are used to generate an Abstract Syntax Tree (AST). Based on this, so-called type hints are extracted and used as input for the model, which consists of two Long Short-Term Memories (LSTM) and a dense layer. The first type hint is the name of the variable. Furthermore, the second type hint is obtained from the code context of the variable. For the third type hint (visible type hint) the data types present in the source code file are analyzed and encoded into a vector. Only the 1024 most frequent data types given in the training dataset are considered. The first two type hints are encoded using Word2Vec \cite{Word2Vec}, which is a static embedding learned on the training data. A drawback of this method is that only words that are present in the training set can be embedded. We investigate the impact of the out-of-vocabulary words in Section \ref{subsec:rq5}.

Afterward, the embedded vectors are taken as input for two separate LSTMs and the output is concatenated with the visible type hints. Next, the feature vector is processed by a fully connected layer and then used for a k-nearest neighbor search \cite{KNN} in the type cluster. Thus, it is possible to predict all data types from the training dataset. To train the system, deep similarity learning is performed using a triplet loss \cite{TripletLoss} function $L$ defined as follows:
\begin{equation}
	\label{equ:TripletLoss}
	L(t_a,t_p,t_n) = \max(0,m+||t_a-t_p||-||t_a-t_n||)
\end{equation}
with a positive scalar margin $m$. To measure the distances between the samples the Euclidean metric is used. The goal of $L$ is to move similar samples closer together ($t_a$ \& $t_p$) and different samples further apart ($t_a$ \& $t_n$) in the cluster.

\section{CrossDomainTypes4Py}
\label{sec:dataset}

This section addresses the creation of our dataset and used methods. The first part explains how we select our dataset domains and find corresponding repositories on the platforms GitHub\footnote{\url{https://github.com}} and Libraries\footnote{\url{https://libraries.io}}. Afterward, we discuss the applied preprocessing steps.
\subsection{Domain Selection}

Code domains can be defined at varying granularity, for example, projects, developers, categories of the software (e.g. embedded, web, scientific calculation), companies, etc. For our dataset, we focus on the category of software (application areas) as a domain, since we expect differences between code from different application domains with respect to structure, programming patterns, used libraries, and also data types. Additionally, there is sufficient data available in public repositories to train and test a machine learning-based system (see Table \ref{tab:Dataset}).

The domains are chosen based on a survey with more than 23,000 Python developers and enthusiasts conducted by JetBrains and the Python Software Foundation \cite{JetBrainsSurvey}. According to this, Python is most commonly used for web development and data analysis. The most utilized libraries in these domains are Flask (web framework) \cite{grinberg2018flask} and NumPy (fundamental package for scientific computing) \cite{Numpy}, respectively. Hence, for our research, we select the web domain (web) with the library Flask and the library NumPy which is generally used for the domain of scientific calculation (cal). These libraries are used to find dependent repositories which belong to one of those domains (see Section \ref{subsec:miningRepo}).

We publicly provide the scripts and tools to generate domain-specific datasets to foster research and researchers for other domains besides the two domains investigated in this paper.

\subsection{Mining Repositories}
\label{subsec:miningRepo}

For mining the repositories, we choose the platforms GitHub and Libraries, on which we search for repositories that depend on the static type-checking tool Mypy\footnote{\url{https://github.com/python/mypy}}. The intention is to ensure that optional type annotations are present in at least a part of the repository (see Section \ref{sec:limitations}). We extend this procedure and check also for dependencies to the libraries Flask and NumPy, in order to be able to assign the repository to a domain.

Since the platforms do not support searching for multiple dependencies at the same time, so we utilize the method explained in the following paragraph. First, we search separately for repositories with dependencies to the three frameworks. For mining the platform Libraries, we consume its API\footnote{\url{https://libraries.io/api/pypi/<Framework>/dependent_repositories}} and query the frameworks separately.

The GitHub API offers no suitable way to query for dependent repositories. Hence, we use web scraping to extract the dependency graph from the website\footnote{\url{https://github.com/<Username>/<Framework>/network/dependents}}. All queries for both platforms are executed automatically. The resulting repositories are then stored in temporary lists. We limit the search to 50,000 repositories per framework (see Section \ref{sec:limitations}). Afterward, these lists are sorted by repository stars and can be filtered if required. The stars are an indicator of popularity and can reflect a tendency about the quality of the repository \cite{stars, munaiah2017curating}.

The temporary lists from both platforms are merged and then used to determine intersections between NumPy \& Mypy and Flask \& Mypy. If repositories have dependencies on all three frameworks, they are included in both subsets, because they are removed during the preprocessing depending on the task (see Section \ref{subsec:preprocessing}). The resulting lists are the basis of the dataset.

The published dataset includes the links to the repositories and a commit hash in order to keep the dataset reproducible. Two example entries are shown in Table \ref{tab:dslist}.

\subsection{Preprocessing Steps}
\label{subsec:preprocessing}

In this section, we discuss the preprocessing steps to make the dataset usable for the Type4Py type inference system. We use the ManyTypes4Py\footnote{\url{https://github.com/saltudelft/many-types-4-py-dataset}} pipeline as a base and adapt it where necessary for our cross-domain setup.\\

\subsubsection{Deduplication}
An essential preprocessing step is to remove duplicates from the dataset, as this harms the performance of machine learning systems \cite{CodeDuplicate}. In particular, for our cross-domain setup, we have to control code duplicates additionally across the datasets. In the first step, we create a list of repositories, which are present in both datasets and randomly remove one-half from one dataset and the other half from the other dataset. The resulting repository lists of the datasets are disjoint.

In the second step, we apply the tool CD4Py\footnote{\url{https://github.com/saltudelft/CD4Py}} to detect file-level duplicates. This tool is also used in the ManyTypes4Py pipeline. It creates a vector representation using the Term Frequency-Inverse Document Frequency (TF-IDF) method to convert the tokenized identifiers of the source code files. The outputs are clusters of duplicates by performing a k-nearest neighbor search. From each cluster, we randomly select one file to remain in the dataset, all others are deleted.\\

\subsubsection{Dataset Split}
The two subsets are randomly split into training, validation, and testing with 70, 10, and 20 percent, respectively. We deviate at this point from the ManyTypes4Py approach and split on project-level rather than on file-level. In the area of type inference splitting on file-level is widely used \cite{Type4py, TypeWriter, Dltpy}, but projects may be split into training and test set. This can lead to leakage of information into the test set, also known as group leakage \cite{kaufman2012leakage}. Furthermore, by splitting on file-level, project-specific data types can be distributed across training and test set, resulting in a higher number of predictable data types, which is not the case in a realistic scenario.

These two problems lead to an overestimation of the performance of the system when the goal is to perform cross-project or more general cross-domain prediction and therefore we conduct the split on project-level.\\

\subsubsection{Feature Extraction}

\begin{table}[t]
	\centering
	\caption{The fields of the JSON file extracted by LibSA4Py.}
	\label{tab:libsa4pstruc}
	\begin{tabular}{ll}
		\hline
		Name of the field     & Description                                    \\ \hline
		\multicolumn{2}{l}{\textbf{Project-Object}}                            \\
		author \& repository  & Author and project name on GitHub              \\
		src\_files            & Path of the project's source code files        \\
		file\_path            & Path of the source code file                   \\ \hline
		\multicolumn{2}{l}{\textbf{Module-Object}}                             \\
		untyped\_seq          & Normalized seq2seq representation              \\
		typed\_seq            & Type of identifiers in untyped\_seq            \\
		imports               & Name of imports                                \\
		variables             & Name and type of variables                     \\
		classes               & Classes of the module (JSON class object)      \\
		funcs                 & Functions of the module (JSON func object)     \\
		set                   & Set of the file (train, valid, test)           \\ \hline
		\multicolumn{2}{l}{\textbf{Class-Object}}                              \\
		name                  & Class name                                     \\
		variables             & Class variables and corresponding type         \\
		funcs                 & Functions of the class (JSON func object)      \\ \hline
		\multicolumn{2}{l}{\textbf{Function-Object}}                           \\
		name                  & Function name                                  \\
		params                & Parameter name and corresponding type          \\
		ret\_exprs            & Return expression                              \\
		ret\_type             & Return type                                    \\
		variables             & Local variables and corresponding type         \\
		params\_occur         & Parameters and their usage in the function     \\
		docstring             & Docstring (with the following three subfields) \\
		docstring.func        & One-line function description                  \\
		docstring.ret         & Description of what the function returns       \\
		docstring.long\_descr & Long description                               \\ \hline
	\end{tabular}
\end{table}

For further preprocessing we take advantage of the LibSA4Py library\footnote{\url{https://github.com/saltudelft/libsa4py}}. It parses the source code and extracts features of interest for machine learning-based type inference systems. The extracted fields and a corresponding description are given in Table \ref{tab:libsa4pstruc}. For more detailed information we refer to \cite{mt4py2021}.\\

\subsubsection{Feature Preparation}
For the preparation of the features, we follow previous works \cite{Type4py,Typilus,TypeWriter}. We remove trivial functions like \code{\_\_len\_\_} with straightforward return types. Furthermore, we exclude the data types \code{Any} and \code{None} because they are not helpful to predict. Moreover, we resolved type aliasing to make the same data types consistent, for example, \code{[]} to \code{List}. In order to reduce the number of different data types, we make a simplification and limit the nested level of data types to two, as Type4Py \cite{Type4py} does. For example \code{List[List[Set[int]]]} is rewritten to \code{List[List[Any]]}. In general, we use fully qualified names for our type annotations to make them consistent across the dataset.

In Table \ref{tab:detailed_overview_mt4p_fl_np} the final amount of samples in all datasets are shown. We see that number of samples of the ManyTypes4Py dataset is in between our two domains web and scientific calculation.

Limitations of our process and the resulting dataset are discussed in Section~\ref{sec:limitations}.

\section{Experimental Results and Evaluation}
\label{sec:evaluation}
This section starts with details about the experiment setup and a description of the evaluation process. In the following our research questions will be motivated, raised, answered, and discussed:

\begin{enumerate}
	\item Are there differences in the distribution of data	types between the domains?
	\item Is the performance of the system similar when evaluated across domains to that which is observed when tested on the training domain?
	\item How do the results change when the evaluation is conducted using only data types known to the system?
	\item How well can the Type4Py method handle class imbalances, and what influence do they have on the results?
	\item What is the impact of the out-of-vocabulary problem on system performance?
	\item What dataset shifts are present, and how can they be mitigated?
\end{enumerate}

\subsection{Experiment and Evaluation Setup}
\label{subsec:exp_eval_setup}
\begin{table*}[tbh]
	\centering
	\caption{A detailed overview of the characteristics of all datasets. The common types refer to data types that occur more than 100 times in the dataset.}
	\label{tab:detailed_overview_mt4p_fl_np}
	\begin{tabular}{@{}rrrrrrrrrrrrr@{}}
		\toprule
		\multirow{2}{*}{Characteristics} & \multicolumn{4}{c}{Web} & \multicolumn{4}{c}{Scientific Calculation} & \multicolumn{4}{c}{ManyTypes4Py}                                                                                       \\ \cmidrule(l){2-13}
		                                 & All                     & Train                                      & Val                              & Test   & All     & Train   & Val    & Test    & All     & Train   & Val    & Test   \\ \midrule
		\textbf{Samples}                 & 341,029                 & 251,064                                    & 27,987                           & 61,978 & 682,354 & 476,768 & 56,854 & 148,732 & 532,522 & 398,152 & 46,577 & 87,793 \\
		Common Samples                   & 240,074                 & 179,877                                    & 20,639                           & 39,558 & 493,813 & 347,520 & 42,786 & 103,507 & 363,553 & 274,200 & 34,420 & 54,933 \\
		Rare Samples                     & 100,955                 & 71,187                                     & 7,348                            & 22,420 & 188,541 & 129,248 & 14,068 & 45,225  & 168,969 & 123,952 & 12,157 & 32,860 \\ \midrule
		\textbf{Unique Types}            & 15,177                  & 7,588                                      & 1,195                            & 8,475  & 27,611  & 14,973  & 2,218  & 14,960  & 24,565  & 13,803  & 1,820  & 11,271 \\
		Common Types                     & 242                     & 232                                        & 158                              & 192    & 381     & 363     & 252    & 332     & 302     & 286     & 168    & 236    \\
		Rare Types                       & 14,935                  & 7,356                                      & 1,037                            & 8,283  & 27,230  & 14,610  & 1966   & 14,628  & 24,263  & 13,517  & 1,652  & 11,035 \\ \bottomrule
	\end{tabular}
\end{table*}
To perform the experiments, we take the available implementation of Type4Py as a template and extend it to our cross-domain setup. We utilize Python 3.6 and the deep learning framework PyTorch. In order to determine the hyperparameters, we conduct a grid search and reuse the configuration for all experiments. We train for 30 epochs and use adam as an optimizer with a learning rate of 0.002 and a batch size of 2,536. The complete configuration can be found in our public repository. For the experiments, an NVIDIA Tesla V100 GPU and an Intel Xeon Platinum 8260 are used.

To perform cross-domain experiments, we train the system on one domain and evaluate it on another domain. To have a comparison of what results can ideally be achieved, we perform a second experiment using only the latter domain, both for training and evaluation. Our two main setups are:
\begin{enumerate}
	\item Setup: \textbf{Web2Cal}
	      \begin{enumerate}
		      \item Training on web domain and evaluation on scientific calculation domain (cal)
		      \item Training and evaluation on scientific calculation domain
	      \end{enumerate}
	\item Setup: \textbf{M4p2Cal}
	      \begin{enumerate}
		      \item Training on ManyTypes4Py (m4p) and evaluation on scientific calculation domain
		      \item Training and evaluation on scientific calculation domain
	      \end{enumerate}
\end{enumerate}

Note that setups 1.b and 2.b are not identical. They use different datasets because of the deduplication step in the preprocessing (see Section \ref{subsec:preprocessing}). The first setup Web2Cal investigates the generalizability of the system from one software domain to another unseen one. In contrast, the second setup M4p2Cal is expected to be an easier task, as the ManyTypes4Py dataset which contains various domains, is used for training and a specific domain for evaluation. This also corresponds to a realistic application scenario. For example, the system should be used in a company with different departments working in various fields. They likely use a pretrained system that is not fine-tuned for the specific fields of the departments. Hence, it is interesting to know for the company how well the system can generalize and what performance could be expected.

We focus on the Web2Cal and M4p2Cal setups for our detailed evaluation. For Cal2Web and Cal2M4p, we omit metrics since the individual results differ only slightly from their inverted counterparts. Still, the conclusions in the following are valid for both directions.

For the evaluation, the top-1 F1-score weighted by the number of samples is used.
Thus, the influence of the more frequent data types on the result is magnified. We further report the accuracy.
All experiments are executed three times to enable a useful significance test and confirm the soundness of our results. To determine whether two results differ significantly, we apply Student's t-test \cite{student1908probable} with a p-value threshold of 0.05. In the tables, the mean and standard deviation of the results in percent are reported.

\subsection{Research Questions and Results}

\textbf{RQ 1: Are there differences in the distribution of data types between the domains?}
\label{subsec:rq1}
\begin{figure}[t]
	\centering
	\includegraphics[width=\linewidth]{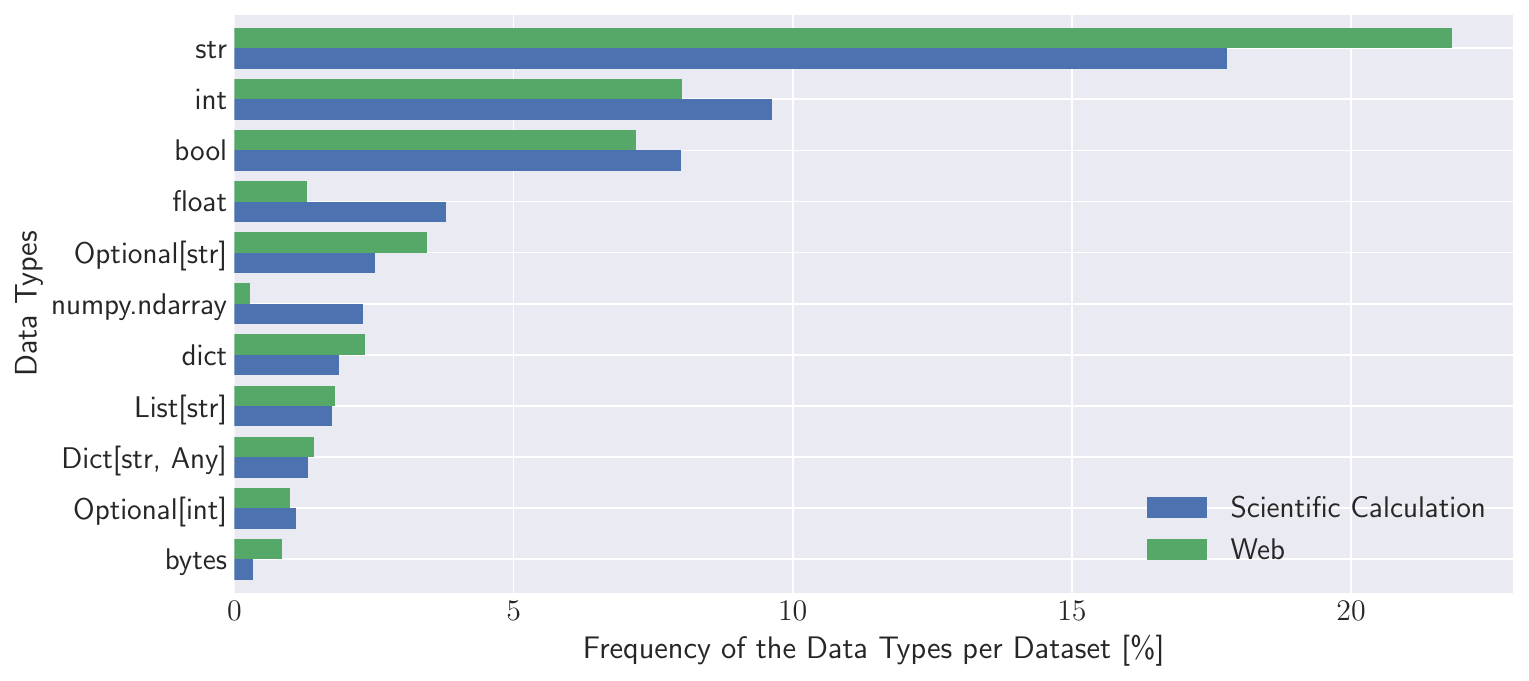}
	\caption{The chart shows the ten most common data types from the web and the scientific calculation domain with their frequency. The trivial data types \textit{None} and \textit{Any} are omitted, because they are not predicted later by the type inference systems.}
	\label{fig:Diagram}
\end{figure}

In the first research question, we analyze the class distribution of the datasets and check if there are differences between the domains. Differences in the distribution indicate a dataset shift that can affect the accuracy of the type inference system (see Section \ref{subsec:dataset_shift}).

In order to answer the research question, we explore the ten most frequent data types from the web and scientific calculation set in Figure~\ref{fig:Diagram}. The three most common data types are built-in data types and are equal for both subsets. As expected, it is noticeable that the data types needed for calculations, such as \textit{bool}, \textit{int}, \textit{float}, and \textit{numpy.ndarray}, occur much more frequently in the scientific calculation subset. In the web subset, on the other hand, the data types \textit{string}, \textit{Optional[str]}, and \textit{dict} are used more often.

We can see the different usage of the data types as well when we compare the list of visible type hints containing the 1,024 most frequent data types from the different domains (see Section \ref{subsec:type4py}). For example, ManyTypes4Py and the scientific calculation domain share only 502 out of 1,024 data types.

When considering the whole datasets, Table \ref{tab:detailed_overview_mt4p_fl_np} shows for instance that the web and scientific calculation domain share only 3,755 classes out of 15,177 and 27,611, respectively. The reason for this is the long-tailed data type distribution with a lot of less frequent data types which are likely to be project- or domain-specific. The aforementioned issue can be observed in the M4p2Cal setup between ManyTypes4Py and the scientific calculation set as well.

We can conclude from our findings that the data types are used differently across the domains and that the distribution of the data types differs. Thus, it can be argued that there is a dataset shift. The following research question examines how this influences the type inference system's results.

\vspace{5pt}
\noindent\fbox{
	\parbox{0.95\linewidth}{
		{Answer to RQ 1: Yes, there is a difference in the distribution of data types which indicates a dataset shift.}
	}
}
\vspace{5pt}

\textbf{RQ 2: Is the performance of the system similar when evaluated across domains to that which is observed when tested on the training domain?}
\begin{table}[t]
	\centering
	\caption{The results of the cross-domain experiments with both setups are shown. The mean of the F1-score, the standard deviation, and the average accuracy in percent are reported.}
	\label{tab:ds_shift_fl_np_mt4p}
	\begin{tabular}{|ll|ll|}
		\hline
		\multicolumn{2}{|l|}{\multirow{2}{*}{\diagbox{Train Set}{Eval Set}}} & \multicolumn{2}{c|}{Cal}                                                                                 \\ \cline{3-4}
		\multicolumn{2}{|l|}{}                                               & \multicolumn{1}{l|}{All Types} & Known Types                                                             \\ \hline
		\multicolumn{1}{|l|}{\multirow{2}{*}{Setup 1}}                       & Web                            & \multicolumn{1}{l|}{49.06 $\pm$ 0.13 (51.6)} & 66.05 $\pm$ 0.17 (69.5)  \\ \cline{2-4}
		\multicolumn{1}{|l|}{}                                               & Cal                            & \multicolumn{1}{l|}{55.27 $\pm$ 0.07 (58.2)} & 69.98 $\pm$ 0,11 (73.63) \\ \hline
		\multicolumn{1}{|l|}{\multirow{2}{*}{Setup 2}}                       & M4p                            & \multicolumn{1}{l|}{45.19 $\pm$ 0.01 (48.1)} & 62.29 $\pm$ 0.02 (66.4)  \\ \cline{2-4}
		\multicolumn{1}{|l|}{}                                               & Cal                            & \multicolumn{1}{l|}{59.34 $\pm$ 0.06 (62.7)} & 72.97 $\pm$ 0.13 (76.9)  \\ \hline
	\end{tabular}
\end{table}
\begin{table*}[t]
	\centering
	\caption{A detailed evaluation of the cross-domain experiments for both setups is shown. The mean of the F1-score, the standard deviation, and the average accuracy in percent are reported.}
	\label{tab:common_rare_types}
	\begin{tabular}{|ll|llll|}
		\hline
		\multicolumn{2}{|r|}{\multirow{3}{*}{\diagbox{Train Set~}{Eval Set}}} & \multicolumn{4}{c|}{Cal}                                                                                                                                                                              \\ \cline{3-6}
		\multicolumn{2}{|r|}{}                                                & \multicolumn{2}{l|}{All Types} & \multicolumn{2}{l|}{Known Types}                                                                                                                                     \\ \cline{3-6}
		\multicolumn{2}{|r|}{}                                                & \multicolumn{1}{l|}{Common}    & \multicolumn{1}{l|}{Rare}                    & \multicolumn{1}{l|}{Common}                  & Rare                                                                   \\ \hline
		\multicolumn{1}{|l|}{\multirow{2}{*}{Setup 1}}                        & Web                            & \multicolumn{1}{l|}{75.46 $\pm$ 0.04 (74.0)} & \multicolumn{1}{l|}{12.35 $\pm$ 0.18 (13.0)} & \multicolumn{1}{l|}{75.46 $\pm$ 0.04 (74.0)} & 45.52 $\pm$ 0.21 (42.8) \\ \cline{2-6}
		\multicolumn{1}{|l|}{}                                                & Cal                            & \multicolumn{1}{l|}{80.25 $\pm$ 0.02 (78.6)} & \multicolumn{1}{l|}{22.66 $\pm$ 0.11 (23.1)} & \multicolumn{1}{l|}{80.29 $\pm$ 0.03 (78.7)} & 48.80 $\pm$ 0.13 (45.0) \\ \hline
		\multicolumn{1}{|l|}{\multirow{2}{*}{Setup 2}}                        & M4p                            & \multicolumn{1}{l|}{73.23 $\pm$ 0.01 (72.1)} & \multicolumn{1}{l|}{8.32 $\pm$ 0.02 (8.2)}   & \multicolumn{1}{l|}{73.23 $\pm$ 0.01 (72.1)} & 33.92 $\pm$ 0.03 (30.6) \\ \cline{2-6}
		\multicolumn{1}{|l|}{}                                                & Cal                            & \multicolumn{1}{l|}{82.55 $\pm$ 0.04 (81.1)} & \multicolumn{1}{l|}{31.40 $\pm$ 0.06 (32.0)} & \multicolumn{1}{l|}{82.55 $\pm$ 0.05 (81.1)} & 55.02 $\pm$ 0.15 (50.0) \\ \hline
	\end{tabular}
\end{table*}

We use our setups defined in Section~\ref{subsec:exp_eval_setup} to answer this question. The results of the experiments are shown in Table~\ref{tab:ds_shift_fl_np_mt4p}. Using the Web2Cal setup 1.a, we measured an F1-score of 49.06\% in comparison to setup 1.b with an F1-score of 55.27\%, which is significantly higher according to the Student's t-test. Consequently, the system has problems generalizing from one specific domain to another.

A more realistic scenario is addressed by our second setup M4p2Cal. We expect a better generalization ability due to the domain diversity in the ManyTypes4Py dataset. However, the results in Table~\ref{tab:ds_shift_fl_np_mt4p} do not confirm our assumption. Setup 2.a achieves an F1-score of 45.19\% and, in comparison, setup 2.b achieves 59.34\%. We observe significantly worse results when the system is used on a domain on which it is not trained. We assume the problems are due to a dataset shift, which is introduced by our domain-specific datasets. When using the system on another than the training domain a decreased performance must be expected. In the following research questions, we analyze the problem in more detail.

\vspace{5pt}
\noindent\fbox{
	\parbox{0.95\linewidth}{
		{Answer to RQ 2: No, when evaluating on another domain, the F1-score decreases by up to 14.15 percentage points compared to training on the corresponding domain.}
	}
}
\vspace{5pt}

\textbf{RQ 3: How do the results change when the evaluation is conducted using only data types known to the system?}

This question is motivated by an analysis of the test sets. We found that in the second setup M4p2Cal about 88 percent of the data types in the scientific calculation test set are unknown to the system because they are not present in the training set (see Table \ref{tab:detailed_overview_mt4p_fl_np}). Thus, about 27~percent of the samples cannot be predicted at all, which affects the performance of the system. The same patterns are confirmed in our first setup Web2Cal. Furthermore, we found that this is also the case within datasets, for example in the ManyTypes4Py dataset only 1,801 out of 11,271 data types from the test set can be predicted (see Table \ref{tab:source-target-classifier}).

For our next experiment, we remove the unknown data types from the test set. This allows us to assess their influence on the result. In M4p2Cal setup 2.a the F1-score increases by 16.99 percentage points to 66.05\%, as well as in setup 2.b, where it increases by 14.71 percentage points to 69.98\% F1-score (see Table \ref{tab:ds_shift_fl_np_mt4p}). These results differ significantly according to the Student's t-test. We observe similar results in our Web2Cal setup 1.a and 1.b. Thus, we conclude that the unknown data types have a great impact on the results and the problem should be addressed. As possible solutions, we propose to use methods from zero-shot learning \cite{zeroshot} or novelty detection \cite{pimentel2014review} and consider the human-in-the-loop for a life-long learning process~\cite{lifelonglearning}.

At the same time, we note that the unknown data types do not fully explain the gap between the cross-domain evaluation (setup a) and the training on the corresponding domain (setup~b).

\vspace{5pt}
\noindent\fbox{
	\parbox{0.95\linewidth}{
		{Answer to RQ 3: F1-score can be significantly improved by up to 16.99 percentage points when removing data types unknown to the system.}
	}
}
\vspace{5pt}

\textbf{RQ 4: How well can the Type4Py method handle class imbalances, and what influence do they have on the results?}

The goal of this research question is to find out how Type4Py deals with class imbalances since deep-learning methods have a significantly worse recognition rate on rarer classes. It is important to analyze this effect in order to better estimate the accuracy of the results of the type inference system in practical applications and to reveal a possible potential for improvement of the method.

We have divided the data types into two groups based on their frequency to address this question. The first group we call \textbf{rare data types}. It contains data types that occur less than 100 times in the dataset. We adopt this threshold of 100 from Mir et al.~\cite{Type4py}. All other data types belong to the group of \textbf{common data types}. This is done separately for every dataset. If we consider the distribution of common and rare data types in the datasets, we notice that there are few common data types with many examples and a lot of rare data types with few examples (see Table \ref{tab:detailed_overview_mt4p_fl_np}). Examples of common data types for the web and scientific calculation set can be seen in Figure~\ref{fig:Diagram}. Rare data types are mostly user-defined or nested data types, which are application-specific.

Table \ref{tab:detailed_overview_mt4p_fl_np} and Figure \ref{fig:Diagram} provide evidence that the distribution of the data types is long-tailed \cite{reed2001pareto}. For instance, the scientific calculation test set consists of 332 common and 14,960 rare data types. It can be assumed that common data types are predicted much better than rare data types because they are predominant during the learning process.

For our experiment, we keep all data types in the training set and evaluate the experiment from RQ 3 according to common and rare data types, illustrated in Table \ref{tab:common_rare_types}.

The rare data types can be predicted significantly worse than the common data types for both setups. If we then evaluate only the data types known by the system, we see that only the result of the rare data types improves significantly, since the unknown data types consist of 99 percent rare data types. The results of the common data types stay the same because they consist mostly of known data types. Nevertheless, the performance of the system is still much better on the common than on the rare data types.
\lstset{language=Python, basicstyle=\tiny,
	frame = single
}
\begin{lstlisting}[caption={Example method from the ManyTypes4Py dataset},captionpos=b, label={lst:code}]
def __init__ (self,
	all_tables: {1} = None,
	tables_with_strings: {2} = None,
	database_directory: {3} = None):

	self.all_tables = all_tables
	self.tables_with_strings = tables_with_strings
	if database_directory:
		self.database_directory = database_directory
		self.connection = sqlite3.connect(database_directory)
		self.cursor = self.connection.cursor()
		self.grammar_str = self.initialize_grammar_str()
		self.grammar = Grammar(self.grammar_str)
		self.valid_actions = self.initialize_valid_actions()
\end{lstlisting}
Ground truth label and prediction:
\begin{enumerate}
	\item Label: \code{Dict[str, List[str]]}\\ Prediction: \code{List[str]}
	\item Label: \code{Dict[str, List[str]]}\\ Prediction: \code{Optional[str]}
	\item Label: \code{str}\\ Prediction: \code{str}
\end{enumerate}

In Listing \ref{lst:code}, we see an example function from the ManyTypes4Py dataset. For simplification, we report only the three arguments of the function predicted by the Type4Py system. In this qualitative example, it is easy for the system to predict the common type string but complicated to predict nested data types. This is in line with our quantitative results.

We summarize that it is important to work on the problem with the rare data types to achieve better results with the system and that the gap in the results between setups a and b is independent of the data type occurrence frequency.

\vspace{5pt}
\noindent\fbox{
	\parbox{0.95\linewidth}{
		{Answer to RQ 4: The F1-score decreases up to 64.91 percentage points for data types that occur less than 100 times in the dataset compared to common data types. When removing the unknown data types, the effect is still the same, but the gap in the results between the common and rare data types is smaller.}
	}
}
\vspace{5pt}

\textbf{RQ 5: What is the impact of the out-of-vocabulary problem on system performance?}
\label{subsec:rq5}

\begin{table}[]
	\centering
	\caption{Results of the out-of-vocabulary experiment, where we use different sets to train the Word2Vec (W2V) model and report the corresponding F1-score in percent after training the Type4Py system with the different embedded datasets. In addition, we report the percentage of out-of-vocabulary words (OOV).}
	\label{tab:oov}
	\begin{tabular}{@{}lllll@{}}
		\toprule
		\multirow{2}{*}{\begin{tabular}[c]{@{}l@{}}W2V \\ Train Data\end{tabular}} &
		\multicolumn{2}{l}{Setup 1}                                                &
		\multicolumn{2}{l}{Setup 2}                                                                                  \\ \cmidrule(l){2-5}
		                                                                           & OOV & F1-score & OOV & F1-score \\ \midrule
		Source Train Set                                                           &
		7.6                                                                        &
		\begin{tabular}[c]{@{}l@{}}48.57 $ \pm $  0.08 \end{tabular}               &
		5.6                                                                        &
		\begin{tabular}[c]{@{}l@{}}44.88 $ \pm $  0.05 \end{tabular}                                                 \\
		Both Train Sets                                                            &
		1.8                                                                        &
		\begin{tabular}[c]{@{}l@{}}49.06 $ \pm $  0.13\end{tabular}                &
		1.3                                                                        &
		\begin{tabular}[c]{@{}l@{}}45.19 $ \pm $  0.01\end{tabular}                                                  \\
		All Sets                                                                   &
		0.9                                                                        &
		\begin{tabular}[c]{@{}l@{}}49.25 $ \pm $  0.04\end{tabular}                &
		0.8                                                                        &
		\begin{tabular}[c]{@{}l@{}}45.22 $ \pm $  0.08\end{tabular}                                                  \\ \bottomrule
	\end{tabular}
\end{table}
When embedding source code, the out-of-vocabulary problem plays a major role in many software engineering tasks \cite{karampatsis2020big,hellendoorn2017deep} since user-defined data types, identifiers, and method names make the vocabulary practically infinite. The embedding method Word2Vec, which is used in the Type4Py system, cannot embed unknown words. Thus, vocabulary that does not appear in the training set of the Word2Vec model is not embedded and the information is lost. We assume that in our cross-domain setup this effect is amplified, since domain-specific vocabulary may be used in the domains.

For our experiments, we create three Word2Vec models for each setup, trained with different data. For the first model, we use in setup Web2Cal the training set from the web domain and in setup M4p2Cal the training set from ManyTypes4Py. The second Word2Vec model is trained with the training sets of both domains, which are web and scientific calculation for the first setup Web2Cal and ManyTypes4Py and scientific calculation for the second setup M4p2Cal. In order to train the third model, we do not only use the training sets like in model 2, we utilize all data from both domains.

In the evaluation, we see that in a realistic scenario where we train the embedding only on the training set of one domain, there are 7.6\% in setup Web2Cal and 5.6\% in setup M4p2Cal unknown words on the other domain (see Table \ref{tab:oov}). This is more than double the number of words that cannot be embedded than in the domain Word2Vec is trained. In the configuration where we train on the training data of both domains, the percentage of unembeddable words decreased significantly and has leveled off for both domains. In the last configuration, it drops even further.
However, there remain some unknown words, because words that occur less than three times in the dataset are excluded from the Word2Vec training.

We use different Word2Vec models, trained on different sets of data, to embed the vectors for the Type4Py system and find that the results of the system are not substantially influenced. When we evaluate according to common and rare data types there is also no difference in the results. Thus we can say that the important information is not stored in the domain-specific vocabulary and in general it is not necessary to further investigate or mitigate this issue.

\vspace{5pt}
\noindent\fbox{
	\parbox{0.95\linewidth}{
		{Answer to RQ 5: We have discovered that in the cross-domain setup, there are significantly more words that cannot be embedded. However, this has no substantial effect on the performance of the system.}
	}
}
\vspace{5pt}

\textbf{RQ 6: What dataset shifts are present, and how can they be mitigated?}
\begin{table}[t]
	\centering
	\caption{The table shows the shared data types between Set 1 and Set 2, their corresponding support in the sets, and the F1-score in percent of a classifier, which classifies the features of both sets from which set they come.}
	\label{tab:source-target-classifier}
	\begin{tabular}{@{}lllllll@{}}
		\toprule
		\multirow{2}{*}{Set 1}               & \multirow{2}{*}{Set 2} & \multicolumn{2}{l}{Types} & \multicolumn{2}{l}{Samples} & \multirow{2}{*}{F1}                \\ \cmidrule(lr){3-6}
		                                     &                        & Common                    & Rare                        & Set 1               & Set 2   &    \\ \midrule
		\multicolumn{2}{l}{\textbf{Setup 1}} &                        &                           &                             &                     &              \\
		Web-Train                            & Web-Test               & 185                       & 1,356                       & 193,441             & 43,688  & 72 \\
		Cal-Train                            & Cal-Test               & 322                       & 3,273                       & 384,637             & 119,026 & 62 \\
		Web-Train                            & Cal-Test               & 198                       & 2,193                       & 205,712             & 110,362 & 71 \\
		Cal-Train                            & Web-Test               & 243                       & 1,244                       & 343,293             & 43,738  & 72 \\ \midrule
		\multicolumn{2}{l}{\textbf{Setup 2}} &                        &                           &                             &                     &              \\
		M4p-Train                            & M4p-Test               & 225                       & 1,576                       & 287,874             & 60,540  & 70 \\
		Cal-Train                            & Cal-Test               & 398                       & 5,828                       & 436,554             & 132,218 & 59 \\
		M4p-Train                            & Cal-Test               & 215                       & 2,034                       & 291,431             & 111,993 & 71 \\
		Cal-Train                            & M4p-Test               & 267                       & 1,425                       & 370,843             & 59,146  & 73 \\ \bottomrule
	\end{tabular}
\end{table}
In the context of this research question, we examine the class distribution as well as the distribution of the features to show the presence of dataset shifts. Afterward, we aim to mitigate the negative impacts of these dataset shifts using methods from the field of transfer learning.

In RQ 1 we discuss the dataset shift in terms of the different distribution of classes. Using the ten most common types from the datasets, the differences in the type hints, and the dataset characteristics in general, we observe strong differences in the distribution of the data types across the datasets.

Furthermore, we investigate the feature distribution across the datasets. For this purpose, we take the features processed by the Type4Py system after the last fully connected layer and before it is used by the k-nearest neighbor classifier (see Section~\ref{subsec:type4py}). In order to evaluate the differences, we adopt the approach of Ganin et al. \cite{dann}. We use features from two datasets to learn a simple classifier to assign the features to a dataset. The more accurate the results of the classifier are, the more dissimilar the features are. For our experiments, we use a tree-based classifier~\cite{geurts2006extremely} in combination with 6-fold cross-validation. The results in Table \ref{tab:source-target-classifier} indicate that inside the specific software domain of scientific calculation, the features are harder to distinguish. The features inside the ManyTypes4Py dataset and across the domains are predicted more accurately by the classifier and subsequently contain more information about their dataset or domain from which they come. According to the results, the performance of the Type4Py system inside the ManyTypes4Py dataset should be similar to the cross-domain setup. We can summarize that there is a shift between the datasets as the features and the distribution of the classes differ.

\begin{figure}[t]
	\centering
	\includegraphics[width=\linewidth]{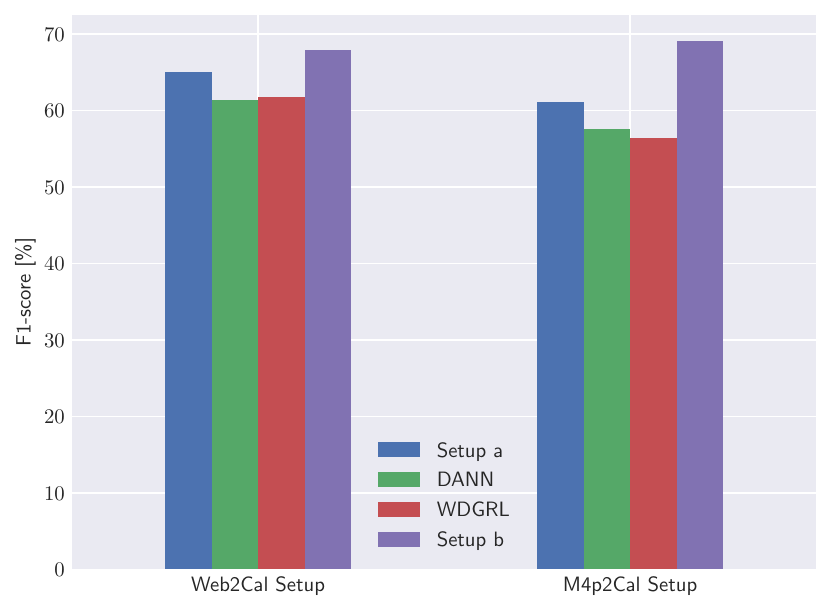}
	\caption{The chart shows the results of the unsupervised domain adaptation methods DANN and WDGRL in comparison to the corresponding setup a and b.}
	\label{fig:res}
\end{figure}

In order to mitigate the dataset shifts, we evaluate two popular methods for unsupervised domain adaptation DANN \cite{dann} and WDGRL \cite{wasserstein} to align the features of the domains in the feature space without the need for additional annotations. They are based on the framework described in \cite{ben2006analysis} and \cite{ben2010theory}. We extend the Type4Py architecture and add the discriminator from the DANN / WDGRL architecture with the proposed hyperparameters. The discriminator is trained together with the Type4Py system and uses the output of the last fully connected layer (see Section~\ref{subsec:type4py}). We observe that these approaches do not provide better results (see Figure~\ref{fig:res}). If we evaluate them according to common and rare data types, we see that the results of both data type groups decreased.

We investigate fine-tuning as an alternative because the unsupervised approaches provide inadequate results. For this purpose, the system is pretrained on a dataset and then learned on the dataset on which it is also evaluated. The drawback of this approach is that we need labeled data from the destination domain but in a real-world scenario labels from the destination domain are often unavailable. When using fine-tuning we can achieve in both setups Web2Cal and M4p2Cal similar results to the model learned directly on the corresponding domain.

\vspace{5pt}
\noindent\fbox{
	\parbox{0.95\linewidth}{
		{Answer to RQ 6: We mitigate the observed shifts in the feature and class distribution with fine-tuning.}
	}
}
\vspace{5pt}

\textbf{Summary}

We experience that when using the Type4Py system on another than the training domain the results decrease by up to an F1-score of 14.15 percentage points in comparison to a training on the corresponding domain. Due to the unbalanced class distribution, the classification accuracy of rare data types is significantly worse than on common data types. The high amount of rare data types also causes a lot of data types that can not be predicted by the system because they are not present in the training set. They decrease the performance by an F1-score up to 16.99 percentage points. Another common issue we investigate is the out-of-vocabulary problem which is present but has no substantial influence on the results of the system. Finally, we show the presence of dataset shifts. In order to mitigate the discovered issues we test different transfer learning methods and find that fine-tuning on the destination domain works best.

\section{Limitations}
\label{sec:limitations}
Our dataset contains only two domains. However, these have been systematically identified through a survey \cite{JetBrainsSurvey} and are the two largest application domains for Python. By providing our tools, an easy extension of our dataset is possible.

While mining our dataset, we limit our search to 50,000 repositories per domain in order to keep the subsets comparable in size to state-of-the-art datasets, e.g. \cite{mt4py2021}.

We use the LibSA4Py library in our preprocessing pipeline for information extraction to maintain comparability with the ManyTypes4Py approach. The library is restricted by its parsing module, which can only handle Python 3, but not the older version Python 2.

The Student's t-test requires a normal distribution of the examined variable, which can be assumed with a sufficiently large sample due to the central limit theorem. Our sample size is limited, which may affect the results of the significance test.

\subsection{Threats to Validity}
Our results on the ManyTypes4Py dataset differ from those presented in the Type4Py paper \cite{Type4py}. However, this does not limit the outcome of this paper because we compare the results from different setups across domains and do not aim to improve the results of the Type4Py paper. The differences in the results are caused by differences in the preprocessing of ManyTypes4Py, which are described in the following. Not all repositories on the dataset list are still available. Additionally, we have to remove duplicates across the datasets.

Furthermore, the data split into training, validation, and test is performed on a project-level because in a realistic scenario, there will not be half of the project in the training and the other half of the project in the test set. Our choice also mitigates the threat of group leakage by projects. This threat is illustrated by results using a file-level split, where the test F1-score in our experiment increases by 7 percentage points compared to using our project-level split.

Except for the experiments around RQ6, we do not apply cross-validation. Instead, we perform each split only once to have consistent test sets throughout our evaluation. We nevertheless repeat each experiment multiple times as stated in Section~\ref{subsec:exp_eval_setup} to account for the effects of random initialization.

While mining our CrossDomainTypes4Py dataset, we increase the number of repositories that contain type annotations by searching for projects that depend on the type checker Mypy. This biases the sampling of the repositories, but is an approved method used by ManyTypes4Py~\cite{mt4py2021} and TypeWriter~\cite{TypeWriter}.

Our process of identifying application domains for Python is based on a single survey~\cite{JetBrainsSurvey}. We selected it because, to our knowledge, it is the largest and most relevant survey of Python developers. It is possible that, had a different study been used, we would have selected other domains. However, the problems we identify in this work are by their nature not specific to certain pairs of domains. Thus, our findings are likely to generalize to further domains.

\section{Conclusion}
\label{sec:conclusion}
We perform the first study of cross-domain generalizability in the field of type inference. We enable this by our publicly available CrossDomainTypes4Py dataset, which consists of two subsets from the domains web and scientific calculation. It contains in total over 1,000,000 type annotations mined on the platforms GitHub and Libraries.

We gain new insights by conducting extensive experiments in various setups. For instance, we observe that the Type4Py system performs significantly worse when doing cross-domain prediction compared to an evaluation on the training domain. Furthermore, we discover a shift between the datasets. In this context, we analyze the differences in the distribution of the data types and the features, which lowers the accuracy of the system results. We apply fine-tuning to mitigate the impact of the dataset shifts. In our investigations, we also show that a large number of out-of-vocabulary words have no substantial impact on the results of the system. Moreover, due to the unbalanced and long-tailed distribution of the dataset, there are many rare data types that the system can only predict with low accuracy.

Based on our findings, we encourage the user of the type inference method to consider the practical environment in which the system is to be deployed. We recommend collecting labeled data of this domain and using it for fine-tuning the system.

Another important aspect that we would like to emphasize is that for a realistic scenario and evaluation of a type inference system, the dataset has to be split into training, validation, and test on a project-level. Splitting it on file-level overestimates the performance of the classifier when we are conducting cross-project or more general cross-domain evaluation.

\subsection*{Future Work}
In this section, we want to give some suggestions for the further development and application of our dataset CrossDomainTypes4Py, as well as possible solutions for the investigated problems regarding the rare data types, dataset shifts, out-of-vocabulary words, and unknown data types.

From a research perspective, the performance of the unpredictable data types should be improved by extending the system to detect them. As a possible solution, we propose methods from the field of novelty detection \cite{pimentel2014review} or zero-shot learning \cite{zeroshot} and consider the human-in-the-loop for a life-long learning process \cite{lifelonglearning}.

To counteract the problem with the rare data types, we recommend using a resampling method like SMOTE \cite{smote} or using importance weighting for the data types during training.

Another aspect for improvement is to replace the static embedding with a contextual embedding to capture more information like TypeBert \cite{TypeBert} does.

Furthermore, it is possible to study how the size of the dataset affects the results of the system.

Besides improving the Type4Py system, it is also possible to further develop our dataset by adding new domains. Moreover, the dataset may also be processed further to enable its usage for related downstream tasks, e.g. code completion \cite{10.1145/3368089.3417058}.

In addition, a descriptive empirical analysis of the repositories and the associated artifacts is also possible, e.g. for empirical analysis of the usage and requirements of type systems in various application domains \cite{10.1145/2989225.2989226}.

\section*{Data Availability}
Our dataset\footnote{\url{https://zenodo.org/record/5747024}} and code\footnote{\url{https://gitlab.com/dlr-dw/type-inference}} are publicly available to ensure reproducibility.

\section*{Acknowledgment}
For our research, we used the High Performance Cluster (HPDA) of the DLR Institute of Data Science, which was funded by the Thuringia Ministry of Economy, Science, and Digital Society. We thank the anonymous reviewers for their valuable feedback and comments.

\bibliographystyle{IEEEtran}
\bibliography{IEEEabrv, type_inference}

\begin{thebibliography}{10}
\providecommand{\url}[1]{#1}
\csname url@samestyle\endcsname
\providecommand{\newblock}{\relax}
\providecommand{\bibinfo}[2]{#2}
\providecommand{\BIBentrySTDinterwordspacing}{\spaceskip=0pt\relax}
\providecommand{\BIBentryALTinterwordstretchfactor}{4}
\providecommand{\BIBentryALTinterwordspacing}{\spaceskip=\fontdimen2\font plus
\BIBentryALTinterwordstretchfactor\fontdimen3\font minus
  \fontdimen4\font\relax}
\providecommand{\BIBforeignlanguage}[2]{{%
\expandafter\ifx\csname l@#1\endcsname\relax
\typeout{** WARNING: IEEEtran.bst: No hyphenation pattern has been}%
\typeout{** loaded for the language `#1'. Using the pattern for}%
\typeout{** the default language instead.}%
\else
\language=\csname l@#1\endcsname
\fi
#2}}
\providecommand{\BIBdecl}{\relax}
\BIBdecl

\bibitem{Pep}
\BIBentryALTinterwordspacing
G.~van Rossum, J.~Lehtosalo, and Łukasz Langa. (2014) Python developer's
  guide: Pep 484 - type hints. [Online]. Available:
  \url{https://www.python.org/dev/peps/pep-0484/}
\BIBentrySTDinterwordspacing

\bibitem{Hanenberg2013AnES}
S.~Hanenberg, S.~Kleinschmager, R.~Robbes, {\'E}.~Tanter, and A.~Stefik, ``An
  empirical study on the impact of static typing on software maintainability,''
  \emph{Empirical Software Engineering}, vol.~19, pp. 1335--1382, 2013.

\bibitem{TypeNotType}
Z.~Gao, C.~Bird, and E.~T. Barr, ``To type or not to type: Quantifying
  detectable bugs in javascript,'' in \emph{2017 IEEE/ACM 39th International
  Conference on Software Engineering (ICSE)}, 2017, pp. 758--769.

\bibitem{10.1145/2989225.2989226}
\BIBentryALTinterwordspacing
T.~S. Heinze, A.~M\o{}ller, and F.~Strocco, ``Type safety analysis for dart,''
  in \emph{Proceedings of the 12th Symposium on Dynamic Languages}, ser. DLS
  2016.\hskip 1em plus 0.5em minus 0.4em\relax New York, NY, USA: Association
  for Computing Machinery, 2016, p. 1–12. [Online]. Available:
  \url{https://doi.org/10.1145/2989225.2989226}
\BIBentrySTDinterwordspacing

\bibitem{Type4py}
\BIBentryALTinterwordspacing
A.~M. Mir, E.~Lato\v{s}kinas, S.~Proksch, and G.~Gousios, ``Type4py: Practical
  deep similarity learning-based type inference for python,'' in
  \emph{Proceedings of the 44th International Conference on Software
  Engineering}, ser. ICSE '22.\hskip 1em plus 0.5em minus 0.4em\relax New York,
  NY, USA: Association for Computing Machinery, 2022, p. 2241–2252. [Online].
  Available: \url{https://doi.org/10.1145/3510003.3510124}
\BIBentrySTDinterwordspacing

\bibitem{TypeBert}
\BIBentryALTinterwordspacing
K.~Jesse, P.~T. Devanbu, and T.~Ahmed, ``Learning type annotation: Is big data
  enough?'' in \emph{Proceedings of the 29th ACM Joint Meeting on European
  Software Engineering Conference and Symposium on the Foundations of Software
  Engineering}, ser. ESEC/FSE 2021.\hskip 1em plus 0.5em minus 0.4em\relax New
  York, NY, USA: Association for Computing Machinery, 2021, p. 1483–1486.
  [Online]. Available: \url{https://doi.org/10.1145/3468264.3473135}
\BIBentrySTDinterwordspacing

\bibitem{TypeWriter}
\BIBentryALTinterwordspacing
M.~Pradel, G.~Gousios, J.~Liu, and S.~Chandra, ``Typewriter: Neural type
  prediction with search-based validation,'' in \emph{Proceedings of the 28th
  ACM Joint Meeting on European Software Engineering Conference and Symposium
  on the Foundations of Software Engineering}, ser. ESEC/FSE 2020.\hskip 1em
  plus 0.5em minus 0.4em\relax New York, NY, USA: Association for Computing
  Machinery, 2020, p. 209–220. [Online]. Available:
  \url{https://doi.org/10.1145/3368089.3409715}
\BIBentrySTDinterwordspacing

\bibitem{Typilus}
\BIBentryALTinterwordspacing
M.~Allamanis, E.~T. Barr, S.~Ducousso, and Z.~Gao, ``Typilus: Neural type
  hints,'' in \emph{Proceedings of the 41st acm sigplan conference on
  programming language design and implementation}, ser. PLDI 2020.\hskip 1em
  plus 0.5em minus 0.4em\relax New York, NY, USA: Association for Computing
  Machinery, 2020, p. 91–105. [Online]. Available:
  \url{https://doi.org/10.1145/3385412.3385997}
\BIBentrySTDinterwordspacing

\bibitem{karampatsis2020big}
R.-M. Karampatsis, H.~Babii, R.~Robbes, C.~Sutton, and A.~Janes, ``Big code!=
  big vocabulary: Open-vocabulary models for source code,'' in \emph{2020
  IEEE/ACM 42nd International Conference on Software Engineering (ICSE)}.\hskip
  1em plus 0.5em minus 0.4em\relax IEEE, 2020, pp. 1073--1085.

\bibitem{hellendoorn2017deep}
V.~J. Hellendoorn and P.~Devanbu, ``Are deep neural networks the best choice
  for modeling source code?'' in \emph{Proceedings of the 2017 11th Joint
  Meeting on Foundations of Software Engineering}, 2017, pp. 763--773.

\bibitem{imbclassvul}
\BIBentryALTinterwordspacing
X.~Ban, S.~Liu, C.~Chen, and C.~Chua, ``A performance evaluation of deep-learnt
  features for software vulnerability detection,'' \emph{Concurrency and
  Computation: Practice and Experience}, vol.~31, no.~19, p. e5103, 2019, e5103
  cpe.5103. [Online]. Available:
  \url{https://onlinelibrary.wiley.com/doi/abs/10.1002/cpe.5103}
\BIBentrySTDinterwordspacing

\bibitem{crossvul}
\BIBentryALTinterwordspacing
X.~Li, Y.~Xin, H.~Zhu, Y.~Yang, and Y.~Chen, ``Cross-domain vulnerability
  detection using graph embedding and domain adaptation,'' \emph{Computers \&
  Security}, vol. 125, p. 103017, 2023. [Online]. Available:
  \url{https://www.sciencedirect.com/science/article/pii/S0167404822004096}
\BIBentrySTDinterwordspacing

\bibitem{JetBrainsSurvey}
\BIBentryALTinterwordspacing
JetBrains. (2022) Python developers survey 2021 results. [Online]. Available:
  \url{https://lp.jetbrains.com/python-developers-survey-2021/}
\BIBentrySTDinterwordspacing

\bibitem{cao2019learning}
K.~Cao, C.~Wei, A.~Gaidon, N.~Arechiga, and T.~Ma, ``Learning imbalanced
  datasets with label-distribution-aware margin loss,'' \emph{Advances in
  neural information processing systems}, vol.~32, 2019.

\bibitem{Word2Vec}
T.~Mikolov, K.~Chen, G.~Corrado, and J.~Dean, ``Efficient estimation of word
  representations in vector space,'' \emph{arXiv preprint arXiv:1301.3781},
  2013.

\bibitem{dann}
Y.~Ganin, E.~Ustinova, H.~Ajakan, P.~Germain, H.~Larochelle, F.~Laviolette,
  M.~Marchand, and V.~Lempitsky, ``Domain-adversarial training of neural
  networks,'' \emph{The journal of machine learning research}, vol.~17, no.~1,
  pp. 2096--2030, 2016.

\bibitem{wasserstein}
J.~Shen, Y.~Qu, W.~Zhang, and Y.~Yu, ``Wasserstein distance guided
  representation learning for domain adaptation,'' in \emph{Proceedings of the
  AAAI Conference on Artificial Intelligence}, vol.~32, no.~1, 2018.

\bibitem{mt4py2021}
A.~M. Mir, E.~Latoskinas, and G.~Gousios, ``Manytypes4py: A benchmark python
  dataset for machine learning-based type inference,'' in \emph{IEEE/ACM 18th
  International Conference on Mining Software Repositories (MSR)}.\hskip 1em
  plus 0.5em minus 0.4em\relax IEEE Computer Society, May 2021, pp. 585--589.

\bibitem{DeepTyper}
\BIBentryALTinterwordspacing
V.~J. Hellendoorn, C.~Bird, E.~T. Barr, and M.~Allamanis, ``Deep learning type
  inference,'' in \emph{Proceedings of the 2018 26th ACM Joint Meeting on
  European Software Engineering Conference and Symposium on the Foundations of
  Software Engineering}, ser. ESEC/FSE 2018.\hskip 1em plus 0.5em minus
  0.4em\relax New York, NY, USA: Association for Computing Machinery, 2018, p.
  152–162. [Online]. Available: \url{https://doi.org/10.1145/3236024.3236051}
\BIBentrySTDinterwordspacing

\bibitem{NL2Type}
R.~S. Malik, J.~Patra, and M.~Pradel, ``Nl2type: Inferring javascript function
  types from natural language information,'' in \emph{2019 IEEE/ACM 41st
  International Conference on Software Engineering (ICSE)}, 2019, pp. 304--315.

\bibitem{LambdaNet}
\BIBentryALTinterwordspacing
J.~Wei, M.~Goyal, G.~Durrett, and I.~Dillig, ``Lambdanet: Probabilistic type
  inference using graph neural networks,'' in \emph{International Conference on
  Learning Representations}, 2020. [Online]. Available:
  \url{https://openreview.net/forum?id=Hkx6hANtwH}
\BIBentrySTDinterwordspacing

\bibitem{Opttyper}
I.~V. Pandi, E.~T. Barr, A.~D. Gordon, and C.~Sutton, ``Opttyper: Probabilistic
  type inference by optimising logical and natural constraints,'' 2021.

\bibitem{Dltpy}
C.~Boone, N.~de~Bruin, A.~Langerak, and F.~Stelmach, ``Dltpy: Deep learning
  type inference of python function signatures using natural language
  context,'' 2019.

\bibitem{Pyinfer}
S.~Cui, G.~Zhao, Z.~Dai, L.~Wang, R.~Huang, and J.~Huang, ``Pyinfer: Deep
  learning semantic type inference for python variables,'' 2021.

\bibitem{BytePairEncoding}
R.~Sennrich, B.~Haddow, and A.~Birch, ``Neural machine translation of rare
  words with subword units,'' \emph{arXiv preprint arXiv:1508.07909}, 2015.

\bibitem{ivanov2021predicting}
V.~Ivanov., V.~Romanov., and G.~Succi., ``Predicting type annotations for
  python using embeddings from graph neural networks,'' in \emph{Proceedings of
  the 23rd International Conference on Enterprise Information Systems - Volume
  1: ICEIS,}, INSTICC.\hskip 1em plus 0.5em minus 0.4em\relax SciTePress, 2021,
  pp. 548--556.

\bibitem{FastText}
P.~Bojanowski, E.~Grave, A.~Joulin, and T.~Mikolov, ``Enriching word vectors
  with subword information,'' \emph{Transactions of the Association for
  Computational Linguistics}, vol.~5, pp. 135--146, 2017.

\bibitem{Hityper}
Y.~Peng, Z.~Li, C.~Gao, B.~Gao, D.~Lo, and M.~Lyu, ``Hityper: A hybrid static
  type inference framework with neural prediction,'' 2021.

\bibitem{Crosslingual}
Z.~Li, X.~Xie, H.~Li, Z.~Xu, Y.~Li, and Y.~Liu, ``Cross-lingual adaptation for
  type inference,'' 2021.

\bibitem{10.1145/3022671.2984041}
\BIBentryALTinterwordspacing
V.~Raychev, P.~Bielik, and M.~Vechev, ``Probabilistic model for code with
  decision trees,'' \emph{SIGPLAN Not.}, vol.~51, no.~10, p. 731–747, oct
  2016. [Online]. Available: \url{https://doi.org/10.1145/3022671.2984041}
\BIBentrySTDinterwordspacing

\bibitem{8816757}
S.~Biswas, M.~J. Islam, Y.~Huang, and H.~Rajan, ``Boa meets python: A boa
  dataset of data science software in python language,'' in \emph{2019 IEEE/ACM
  16th International Conference on Mining Software Repositories (MSR)}, 2019,
  pp. 577--581.

\bibitem{orru2015curated}
M.~Orr{\'u}, E.~Tempero, M.~Marchesi, R.~Tonelli, and G.~Destefanis, ``A
  curated benchmark collection of python systems for empirical studies on
  software engineering,'' in \emph{Proceedings of the 11th International
  Conference on Predictive Models and Data Analytics in Software Engineering},
  2015, pp. 1--4.

\bibitem{finlayson2021clinician}
S.~G. Finlayson, A.~Subbaswamy, K.~Singh, J.~Bowers, A.~Kupke, J.~Zittrain,
  I.~S. Kohane, and S.~Saria, ``The clinician and dataset shift in artificial
  intelligence,'' \emph{The New England journal of medicine}, vol. 385, no.~3,
  p. 283, 2021.

\bibitem{janez2022review}
F.~J{\'a}{\~n}ez-Martino, R.~Alaiz-Rodr{\'\i}guez, V.~Gonz{\'a}lez-Castro,
  E.~Fidalgo, and E.~Alegre, ``A review of spam email detection: analysis of
  spammer strategies and the dataset shift problem,'' \emph{Artificial
  Intelligence Review}, pp. 1--29, 2022.

\bibitem{MorenoTorres2012AUV}
J.~G. Moreno-Torres, T.~Raeder, R.~Ala{\'i}z-Rodr{\'i}guez, N.~Chawla, and
  F.~Herrera, ``A unifying view on dataset shift in classification,''
  \emph{Pattern Recognit.}, vol.~45, pp. 521--530, 2012.

\bibitem{KNN}
T.~Cover and P.~Hart, ``Nearest neighbor pattern classification,'' \emph{IEEE
  Transactions on Information Theory}, vol.~13, no.~1, pp. 21--27, 1967.

\bibitem{TripletLoss}
D.~Cheng, Y.~Gong, S.~Zhou, J.~Wang, and N.~Zheng, ``Person re-identification
  by multi-channel parts-based cnn with improved triplet loss function,'' in
  \emph{2016 IEEE Conference on Computer Vision and Pattern Recognition
  (CVPR)}, 2016, pp. 1335--1344.

\bibitem{grinberg2018flask}
M.~Grinberg, \emph{Flask web development: developing web applications with
  python}.\hskip 1em plus 0.5em minus 0.4em\relax " O'Reilly Media, Inc.",
  2018.

\bibitem{Numpy}
\BIBentryALTinterwordspacing
C.~R. Harris, K.~J. Millman, S.~J. van~der Walt, R.~Gommers, P.~Virtanen,
  D.~Cournapeau, E.~Wieser, J.~Taylor, S.~Berg, N.~J. Smith, R.~Kern, M.~Picus,
  S.~Hoyer, M.~H. van Kerkwijk, M.~Brett, A.~Haldane, J.~F. del R{\'{i}}o,
  M.~Wiebe, P.~Peterson, P.~G{\'{e}}rard-Marchant, K.~Sheppard, T.~Reddy,
  W.~Weckesser, H.~Abbasi, C.~Gohlke, and T.~E. Oliphant, ``Array programming
  with {NumPy},'' \emph{Nature}, vol. 585, no. 7825, pp. 357--362, Sep. 2020.
  [Online]. Available: \url{https://doi.org/10.1038/s41586-020-2649-2}
\BIBentrySTDinterwordspacing

\bibitem{stars}
H.~Borges and M.~T. Valente, ``What’s in a github star? understanding
  repository starring practices in a social coding platform,'' \emph{Journal of
  Systems and Software}, vol. 146, pp. 112--129, 2018.

\bibitem{munaiah2017curating}
N.~Munaiah, S.~Kroh, C.~Cabrey, and M.~Nagappan, ``Curating github for
  engineered software projects,'' \emph{Empirical Software Engineering},
  vol.~22, no.~6, pp. 3219--3253, 2017.

\bibitem{CodeDuplicate}
\BIBentryALTinterwordspacing
M.~Allamanis, ``The adverse effects of code duplication in machine learning
  models of code,'' in \emph{Proceedings of the 2019 ACM SIGPLAN International
  Symposium on New Ideas, New Paradigms, and Reflections on Programming and
  Software}, ser. Onward! 2019.\hskip 1em plus 0.5em minus 0.4em\relax New
  York, NY, USA: Association for Computing Machinery, 2019, p. 143–153.
  [Online]. Available: \url{https://doi.org/10.1145/3359591.3359735}
\BIBentrySTDinterwordspacing

\bibitem{kaufman2012leakage}
S.~Kaufman, S.~Rosset, C.~Perlich, and O.~Stitelman, ``Leakage in data mining:
  Formulation, detection, and avoidance,'' \emph{ACM Transactions on Knowledge
  Discovery from Data (TKDD)}, vol.~6, no.~4, pp. 1--21, 2012.

\bibitem{student1908probable}
Student, ``The probable error of a mean,'' \emph{Biometrika}, pp. 1--25, 1908.

\bibitem{zeroshot}
\BIBentryALTinterwordspacing
W.~Wang, V.~W. Zheng, H.~Yu, and C.~Miao, ``A survey of zero-shot learning:
  Settings, methods, and applications,'' \emph{ACM Trans. Intell. Syst.
  Technol.}, vol.~10, no.~2, jan 2019. [Online]. Available:
  \url{https://doi.org/10.1145/3293318}
\BIBentrySTDinterwordspacing

\bibitem{pimentel2014review}
M.~A. Pimentel, D.~A. Clifton, L.~Clifton, and L.~Tarassenko, ``A review of
  novelty detection,'' \emph{Signal processing}, vol.~99, pp. 215--249, 2014.

\bibitem{lifelonglearning}
\BIBentryALTinterwordspacing
G.~I. Parisi, R.~Kemker, J.~L. Part, C.~Kanan, and S.~Wermter, ``Continual
  lifelong learning with neural networks: A review,'' \emph{Neural Networks},
  vol. 113, pp. 54--71, 2019. [Online]. Available:
  \url{https://www.sciencedirect.com/science/article/pii/S0893608019300231}
\BIBentrySTDinterwordspacing

\bibitem{reed2001pareto}
W.~J. Reed, ``The pareto, zipf and other power laws,'' \emph{Economics
  letters}, vol.~74, no.~1, pp. 15--19, 2001.

\bibitem{geurts2006extremely}
P.~Geurts, D.~Ernst, and L.~Wehenkel, ``Extremely randomized trees,''
  \emph{Machine learning}, vol.~63, no.~1, pp. 3--42, 2006.

\bibitem{ben2006analysis}
S.~Ben-David, J.~Blitzer, K.~Crammer, and F.~Pereira, ``Analysis of
  representations for domain adaptation,'' \emph{Advances in neural information
  processing systems}, vol.~19, 2006.

\bibitem{ben2010theory}
S.~Ben-David, J.~Blitzer, K.~Crammer, A.~Kulesza, F.~Pereira, and J.~W.
  Vaughan, ``A theory of learning from different domains,'' \emph{Machine
  learning}, vol.~79, no.~1, pp. 151--175, 2010.

\bibitem{smote}
N.~V. Chawla, K.~W. Bowyer, L.~O. Hall, and W.~P. Kegelmeyer, ``Smote:
  synthetic minority over-sampling technique,'' \emph{Journal of artificial
  intelligence research}, vol.~16, pp. 321--357, 2002.

\bibitem{10.1145/3368089.3417058}
\BIBentryALTinterwordspacing
A.~Svyatkovskiy, S.~K. Deng, S.~Fu, and N.~Sundaresan, ``Intellicode compose:
  Code generation using transformer,'' in \emph{Proceedings of the 28th ACM
  Joint Meeting on European Software Engineering Conference and Symposium on
  the Foundations of Software Engineering}, ser. ESEC/FSE 2020.\hskip 1em plus
  0.5em minus 0.4em\relax New York, NY, USA: Association for Computing
  Machinery, 2020, p. 1433–1443. [Online]. Available:
  \url{https://doi.org/10.1145/3368089.3417058}
\BIBentrySTDinterwordspacing

\end{thebibliography}
\end{document}